\documentclass[12pt]{article}
\usepackage{graphicx}
\usepackage{soul}
\usepackage{amsmath, amssymb}

\usepackage{microtype}
\usepackage[title]{appendix}
\usepackage[hidelinks]{hyperref} 
\usepackage{comment}

\numberwithin{equation}{section} 
\graphicspath{{images/}}
\begin{document}

\bigskip 

\begin{center}
\textbf{The Spherical-Rindler Framework: From Compact Minkowski\\Regions to Black-Hole and Cosmological Solutions}

\smallskip

\bigskip

Edgar Alejandro León\footnote{ealeon@uas.edu.mx}

\bigskip \

\textit{Facultad de Ciencias Físico-Matemáticas, Universidad Autónoma}

\textit{de Sinaloa, 80010, Culiacán, México.}\bigskip

\end{center}

\bigskip \ 

\begin{abstract}
In this article we first develop novel Rindler-type representations of flat spacetime by demonstrating that the standard hyperbolic transformation is a member of an infinite family of coordinate mappings. We specifically introduce cyclic coordinates, which —in contrast to the conventional Rindler wedge— delineate a compact region of Minkowski spacetime. By extending this framework, and motivated by near-horizon coordinates in Schwarzschild metric, we propose a class of `Spherical-Rindler metrics.' We demonstrate the utility of this approach by deriving and analyzing a black-hole solution and a cosmological metric, both emerging naturally from a Spherical-Rindler origin. Our results highlight unique geometric properties of these solutions, providing new insights into the relationship between accelerated frames and global spacetime curvature.
\end{abstract}

\bigskip

Keywords: General relativity, Cosmology, Exact Solutions.

Pacs numbers: 04.20.Cv, 04.20.Jb, 04.70.Bw, 98.80.-k

January 2026

\newpage

\section{Introduction} \label{Intro}
General relativity is a dynamical theory of spacetime wherein absolutely simultaneity is replaced by a causal structure defined by local light cones. A cornerstone of the theory is general covariance -- the principle that physical laws remain invariant under arbitrary coordinate transformations [1][2]. Nevertheless, a clever choice of a coordinate system is essential for elucidating the physical properties of a manifold, such as its association with specific observer classes, global topological characteristics, and the singular behavior near horizons. This interplay between coordinate-independent physics and coordinate-dependent interpretation has remained a central theme in the study of black holes and FLRW cosmologies since its very inception [3]-[5]. Nevertheless, it continues to drive contemporary research in gravitational physics (see for instance Refs. [6]-[10]).   

To explore this nuances, we begin with the simplest possible arena: flat spacetime. Minkowski spacetime is characterized by the metric%
\begin{equation}
ds^{2}=g_{\mu \nu }dx^{\mu }dx^{\nu }=-dt^{2}+dx^{2}+dy^{2}+dz^{2}. 
\tag{1.1}
\end{equation}%
Throughout the article, Greek indices like $\alpha ,\mu ,...$ are spatiotemporal, ranging from $0$ to $3$, while Latin indices $i,j,...$ vary from $1$ to $3$. In the basis with coordinates $x^{\alpha }=(t,x,y,z)$, the metric components are $g_{\mu\nu}=\eta _{\mu \nu }=diag(-1,1,1,1)$. To change representation in a given patch, we shall be using the tensor transformation%
\begin{equation}
g_{\mu \prime \nu \prime }=\frac{\partial x^{\mu }}{\partial x^{\mu \prime }}%
\frac{\partial x^{\nu }}{\partial x^{\nu \prime }}g_{\mu \nu }.  \tag{1.2}
\end{equation}%
More precisely, we consider the two-dimensional version, which actually is the most common choice for spherically-symmetric spacetimes [10][11].

Throughout the text we assume metric solutions to the standard Einstein's field equations, namely%
\begin{equation}
R_{\mu \nu }-\frac{1}{2}g_{\mu \nu }(R-2\Lambda )=8\pi GT_{\mu \nu }. 
\tag{1.3}
\end{equation}%
On the left hand $R_{\mu \nu }$ is the Ricci tensor and $R$ its contraction $R=g^{\mu \nu }R_{\mu \nu }$, and we also have included the cosmological constant $\Lambda $. On the right hand side, $T_{\mu \nu }$ is the energy-momentum tensor. We shall specify its properties when needed.

The remainder of this article is structured as follows:

Section 2 deals with a `Rindler-type' coordinates, a generalization of the custom Rindler representation [12][13]. In this family of solutions, most can have the usual association with accelerated observers at constant $\rho$. Only two particular cases are of distinct nature: one as an analogue to Tortoise coordinates, and other representing a compact region of Minkowski spacetime. This is the analogue of a recent representation studied for cosmological and black hole solutions [11].

We start Section 3 by obtaining three distinct representations of Schwarzschild's spacetime: near-horizon, Tortoise and Kruskal-Szekeres. We see a simple way to in which the last two are related, one arising from an explicit cartesian (1+1)-conformal transformation, while the other uses the analogue development for Rindler coordinates. Considering Spherical-Rindler metrics, several possibilities arise, and one of these takes precisely the form of near-horizon coordinates. By taking it as a global metric, not just approximation, we contrast its properties with the Schwarzschild case. We obtain relevant quantities such as stable orbits, ISCO, and the effective potential energy associated to this metric. We also mention a curious property for a flat cylindrical embedding, analogue of the Flamm's paraboloid.

In Section 4 we consider two forms where Rindler representations appear in cosmology. The first is Milne Universe, which implies an analogue to Rindler wedge in Minkowski spacetime. The second way Rindler appears is a second form of Spherical-Rindler metric. We contrast some properties with de-Sitter spacetime.

Finally, in Section 5 we make some final comments, and mention possible future open paths in these subjects.

\section{Rindler in Minkowski Spacetime} \label{Minkowski}
Many coordinate representations reveal key physical properties and symmetries. For instance, the cartesian and spherical representations of (1.1) can explicitly yield translational or rotational invariance, respectively. Also, a Lorentz boost can be described by hyperbolic transformations, and its physical significance lies in its role as a change to a new inertial reference frame.

In this work we use Rindler representations to explore transformations related to the physics of various spacetimes. We begin with general transformations commonly associated with Minkowski spacetime, and then generalize the procedure to apply it more broadly.

\subsection{Transforming the Rindler metric} \label{RindlerPrev}

Consider the two-dimensional metric%
\begin{equation}
ds^{2}=-\rho ^{2}dT^{2}+d\rho ^{2}.  \tag{2.1}
\end{equation}%
Note that $T$ and $\rho $ are simply coordinate labels and are not necessarily associated with proper time or any curvilinear coordinate system. Given this form alone, one can not infer wether the four-dimensional version is flat. Nevertheless, it will serve as motivation for the Spherical-Rindler metrics.
In the remainder of this section, we consider the transformation of the first terms of (1.1), that is a four-dimensional flat spacetime of origin.

We denote by $g_{\mu ^{\prime }\nu ^{\prime }}$ the components of the metric (2.1), while the reduced (1+1) version of (1.1) (with $dy=dz=0$) are $g_{\mu \nu }$. The  $g_{0^{\prime}0^{\prime }}$, $g_{1^{\prime }1^{\prime }}$ and $g_{0^{\prime }1^{\prime }}$ components of the transformation (1.2) lead to [10][11]:

\begin{equation}
-\rho ^{2}=-\left(\frac{\partial t}{\partial T}\right) ^{2}+\left( \frac{\partial x}{\partial T}\right) ^{2},  \tag{2.2}
\end{equation}%
\begin{equation}
1=-\left( \frac{\partial t}{\partial \rho }\right) ^{2}+\left(\frac{\partial x}{\partial \rho }\right) ^{2} \tag{2.3}
\end{equation}%
and
\begin{equation}
\frac{\partial t}{\partial T}\frac{\partial t}{\partial \rho }=\frac{\partial x}{\partial T}\frac{\partial x}{\partial \rho }.  \tag{2.4}
\end{equation}

Now extract $(\partial x/\partial T)^{2}$ and $(\partial x/\partial \rho)^{2}$ from (2.2) and (2.3). Multiply and substitute into the square of Eq. (2.4). Canceling terms and using (2.3) again , the simple relation $\partial t/\partial T=\rho \partial x/\partial \rho$ arises. A similar calculation, but now using $(\partial t/\partial T)^{2}$ and $(\partial t/\partial \rho)^{2}$, leads to $\partial x/\partial T=\rho \partial t/\partial \rho$. In turn, these relations allow us to assume $t=f(T)g(\rho )$ and $x=h(T)q(\rho )$. Substitution yields%
\begin{equation}
\frac{1}{h}\frac{df}{dT}=\frac{\rho }{g}\frac{dq}{d\rho }=\alpha ,  \tag{2.5}
\end{equation}%
and%
\begin{equation}
\frac{1}{f}\frac{dh}{dT}=\frac{\rho }{q}\frac{dg}{d\rho }=\beta ,  \tag{2.6}
\end{equation}%
where $\alpha $ and $\beta $ are constants. Furthermore, both relations can be joined in two ways. First, by taking the next derivative. For instance, take $d^{2}f/dT^{2}=\alpha dh/dT$ from (2.5) and then substitute $dh/dT=\beta f$ on the right . This way we obtain%
\begin{equation}
\frac{d^{2}f}{dT^{2}}=\alpha \beta f,  \tag{2.7}
\end{equation}%
as well as a similar relation for $h$, i.e. $d^{2}h/dT^{2}=\alpha \beta h$.
Furthermore, due to the appearance of the factor $\rho $ in (2.5) and (2.6), the analogous equation for $g(\rho )$ is
\begin{equation}
\rho ^{2}\frac{d^{2}g}{d\rho ^{2}}+\rho \frac{dg}{d\rho }=\alpha \beta g, 
\tag{2.8}
\end{equation}%
with an equivalent differential equation for the function $q$.

A second type of relation arises dividing (2.6) by (2.5). Then $\beta fdf-\alpha hdh=0$ appears, with solution

\begin{equation}
\alpha h^{2}-\beta f^{2}=\gamma .  \tag{2.9}
\end{equation}%
The other two functions are also related by

\begin{equation}
\beta q^{2}-\alpha g^{2}=\lambda ,  \tag{2.10}
\end{equation}%
where $\gamma $ and $\lambda $ are constants.

\subsection{A family of Rindler-type transformations} \label{RindlerType}

For our purposes it suffices to consider only three possibilities for $\alpha \beta $: positive, zero or negative. Ref. [11] shows the convenience of choosing this division for different representations of various spacetimes.

\textit{Case $\alpha \beta >1$: Rindler-type solutions.}

Let us define $\eta =\sqrt{\alpha \beta }$ and $w=\ln \rho $, consistent with $\rho >0$. Then we have the identities $dg/d\rho =\left( 1/\rho \right)
\left( dg/dw\right) $ and $d^{2}g/d\rho ^{2}=\left( 1/\rho ^{2}\right)
\left( d^{2}g/dw^{2}-dg/dw\right) $. This turns (2.8) into%
\begin{equation}
\frac{d^{2}g}{dw^{2}}=\eta ^{2}g.  \tag{2.11}
\end{equation}

Since there is an analogous equation for $q$, we can choose as solutions $%
g=e^{\pm \eta w}/\sqrt{\alpha }=\rho ^{\pm \eta }/\sqrt{\alpha }$ and $q=\pm
e^{\pm \eta w}/\sqrt{\beta }=\pm \rho ^{\pm \eta }/\sqrt{\beta }$ (same
sign for both). Then we have $\lambda =0$ in (2.10). Note that a linear combination
does not work.

Now, (2.7) and its analogue for $h$ can be solved by hyperbolic or exponential functions. For instance, take $f=\sqrt{\alpha }\eta ^{-1}\sinh
(\eta T)$ and $h=\sqrt{\beta }\eta ^{-1}\cosh (\eta T)$. This choice is
consistent with (2.5) and (2.6), and also with $\gamma =\eta ^{2}$ in (2.9).
Substituting in $t=fg$ and $x=hq$, we get

\begin{equation}
\begin{array}{c}
t=\eta ^{-1}\rho ^{\pm \eta }\sinh (\eta T) \\ 
\\ 
x=\pm \eta ^{-1}\rho
^{\pm \eta }\cosh (\eta T),%
\end{array}
\tag{2.12}
\end{equation}%
which lead to%
\begin{equation}
-dt^{2}+dx^{2}=\rho ^{-2(1\mp \eta )}\left( -\rho ^{2}dT^{2}+d\rho
^{2}\right) .  \tag{2.13}
\end{equation}

The last two relations mark a family of Rindler-type solutions, parameterized by any $\eta >0$. The textbook Rindler transformation is obtained by choosing $+\eta =1$ in (2.12), which is equivalent to choosing $\rho ^{-2(1\mp \eta )}=1$ in (2.13).

\textit{Case $\alpha \beta =0$. Tortoise-like coordinate.}

Now we assume $\alpha =1$ and $\beta =0$, the opposite being redundant since it just swaps the results. By (2.6), we can set $h$ and $g$ equal to one, and due to (2.5) $f=T$ and $q=\ln \rho $ are suitable choices. Then we have $t=T$ and $x=\ln \rho $. Differentiation leads to $-dt^{2}+dx^{2}=\rho ^{-2}(-\rho ^{2}dT^{2}+d\rho ^{2})$. Note that this is the missing solution in (2.13). It is trivial, since it only redefines the spatial coordinate $\rho =e^{x}$. At $\rho =const.$, we have the world-line of a static observer. Still, for $\rho >0$ and $T\in(-\infty ,\infty )$, it covers the entire (1+1) Minkowski spacetime. It turns out that $x=\ln \rho $ is the analogue of Tortoise coordinate, the most simple non-trivial conformal flat transformation for Schwarzschild's metric [11].

\textit{Case $\alpha \beta <1$: Cyclic coordinates}.

Without loss of generality, we choose $\alpha =1$ and $\beta =-1$. This turns (2.7) and its analog for $h$ into harmonic oscillator type equations. The functions $f=A\sin T$ and $h=A\cos T$ are consistent with (2.5) and (2.6), as well as with $\gamma =A^{2}$ in (2.9). Again, $w=\ln\rho $ leads to $d^{2}g/dw^{2}+g=0$ instead of (2.11). Since $q$ satisfies an analogous relation, it is required that $\lambda =-1$ in (2.10). We choose $g=\cos w$ and $q=\sin w$ as solutions. Since $t=fg$ and $x=hq$, we have the transformation

\begin{equation}
t=A\sin T\cos \left( \ln \rho \right) \text{,\qquad }x=A\cos T\sin \left(\ln \rho \right) .  \tag{2.14}
\end{equation}

Now we have a compact, albeit oscillating, representation. Substituting this into $ds^{2}=-dt^{2}+dx^{2}$, we get

\begin{equation}
ds^{2}=\omega^2 (-\rho ^{2}dT^{2}+d\rho ^{2})  \tag{2.15}
\end{equation}%
where the conformal factor is

\begin{equation}
\omega^{2} =A^{2}\rho ^{-2}\left( \cos ^{2}T\cos ^{2}\ln \rho -\sin ^{2}T\sin^{2}\ln \rho \right) .  \tag{2.16}
\end{equation}%
Also, (2.14) implies%
\begin{equation}
\frac{x^{2}}{\sin ^{2}\left( \ln \rho \right) }+\frac{t^{2}}{\cos ^{2}\left(
\ln \rho \right) }=A^{2}.  \tag{2.17}
\end{equation}

\begin{figure}[ph]
\centerline{\includegraphics[width=2.8in]{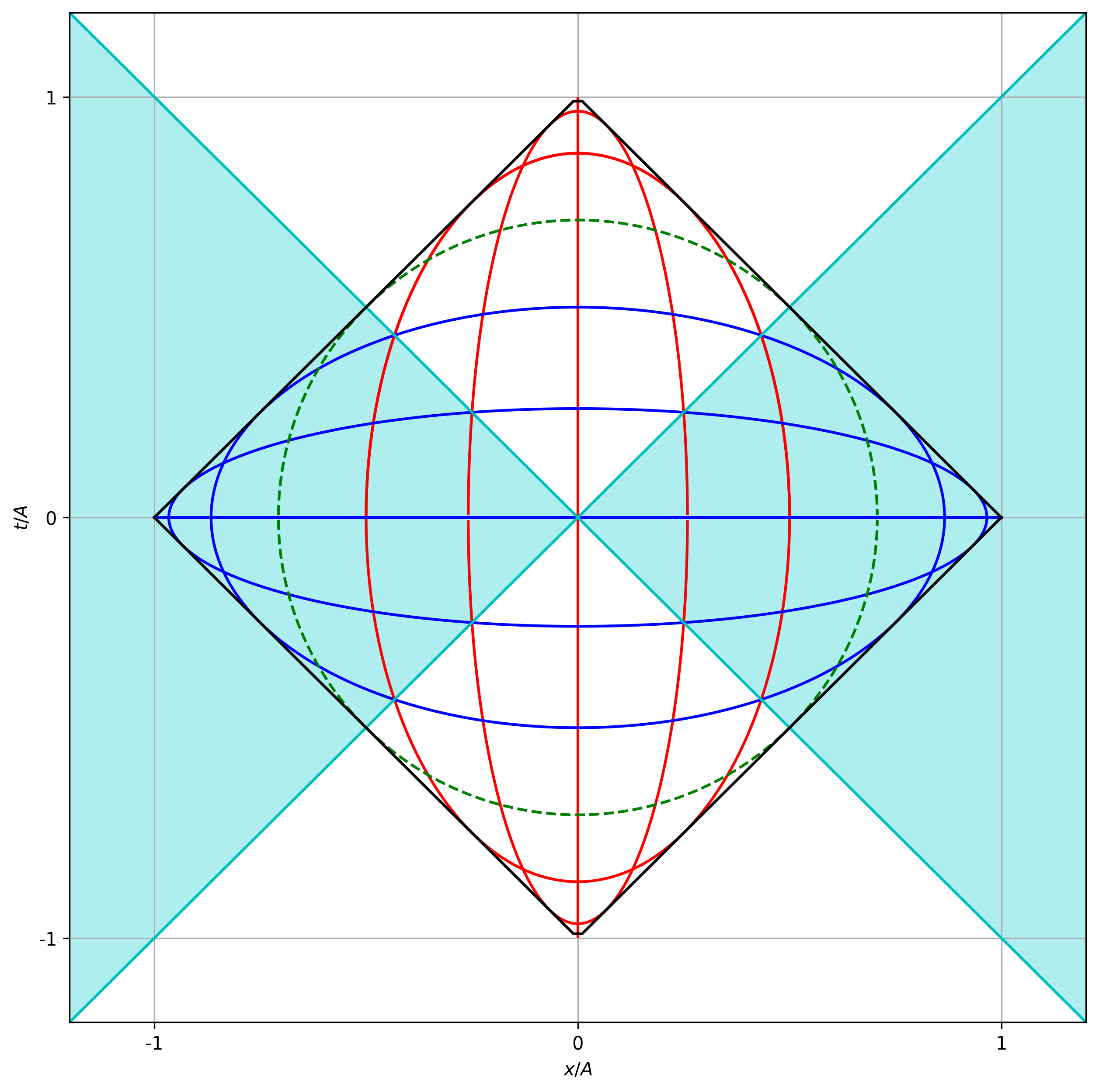}}
\vspace*{8pt}
\caption{Pulsating coordinates described above. The light region is interior of the light-cone respect to the event $(t,x)=(0.0)$.}
\label{cyclic}
\end{figure}

That is, curves with $\rho =const.$ correspond to ellipses in the Minkowski $(t,x)$ representation (see Fig. \ref{cyclic}).

In fact, here $\rho $ is a cyclic coordinate, and a suitable range for it is $1<\rho <\exp (\pi /2)$. Varying $\rho =const.$ initially from its minimum, takes the degenerate vertical ellipse ($t=\pm A,~x=0$), which becomes continuously less elongated, up to $\rho =\exp (\pi /4)$, the value for a circumference of radius $A/\sqrt{2}$. Increasing $\rho $ then leads to more and more elongated horizontal ellipses that approach the degenerate horizontal ellipse ($t=0,~x=\pm A$) as $\rho\rightarrow \exp (\pi /2)$.

Disregarding degenerate ellipses, an accelerated observer can be associated with a portion of the above curves. Since the local coordinate velocity is given by $dx/dt=-\tan ^{2}(\ln \rho )t/x$, the portion with $\left \vert t/x\right \vert <\tan ^{-2}(\ln \rho )$ has a time-like tangent vector. For the circumference $\rho =\exp (\pi /4)$, the limit is the region with $\left\vert x\right \vert >A/2$. The remainder of each curve cannot be associated with a local accelerated observer, since the tangent vectors are space-like.

Clearly, we are covering a compact region of Minkowski spacetime. More specifically, let us call $a^{2}=\sin ^{2}(\ln \rho )$, which turns (2.17) into $x^{2}/a^{2}+t^{2}/(1-a^{2})=A^{2}$. This can be rewritten as
\begin{equation}
A^{2}a^{4}+\left( t^{2}-x^{2}-A^{2}\right) a^{2}+x^{2}=0.  \tag{2.18}
\end{equation}

This is a quadratic equation for $a^{2}$, where the discriminant is $D\equiv \left(t^{2}-x^{2}-A^{2}\right) ^{2}-4A^{2}x^{2}$. We have three possibilities: $D>0$ indicates that two values of $a$, i.e., two distinct ellipses, touch a given point $(t,x)$. $D<0$ indicates the region that is not reached by any ellipse bounded by $A^{2}$. The limit of the Minkowski spacetime region that can be represented is the envelope of $D>0$, which is $D=0$. By simple algebraic calculation, this is equivalent to $t=\pm (x\pm A)$. As can be seen by varying $\rho $, one obtains the interior of a rhombus, whose vertices are $(t,x)=(\pm A,0)$ and $(0,\pm A)$. All of this is represented in Fig. \ref{cyclic}.

\subsection{Accelerated observers and conformal flatness} \label{accelerObs}

\begin{figure}[ph]
\centerline{\includegraphics[width=2.8in]{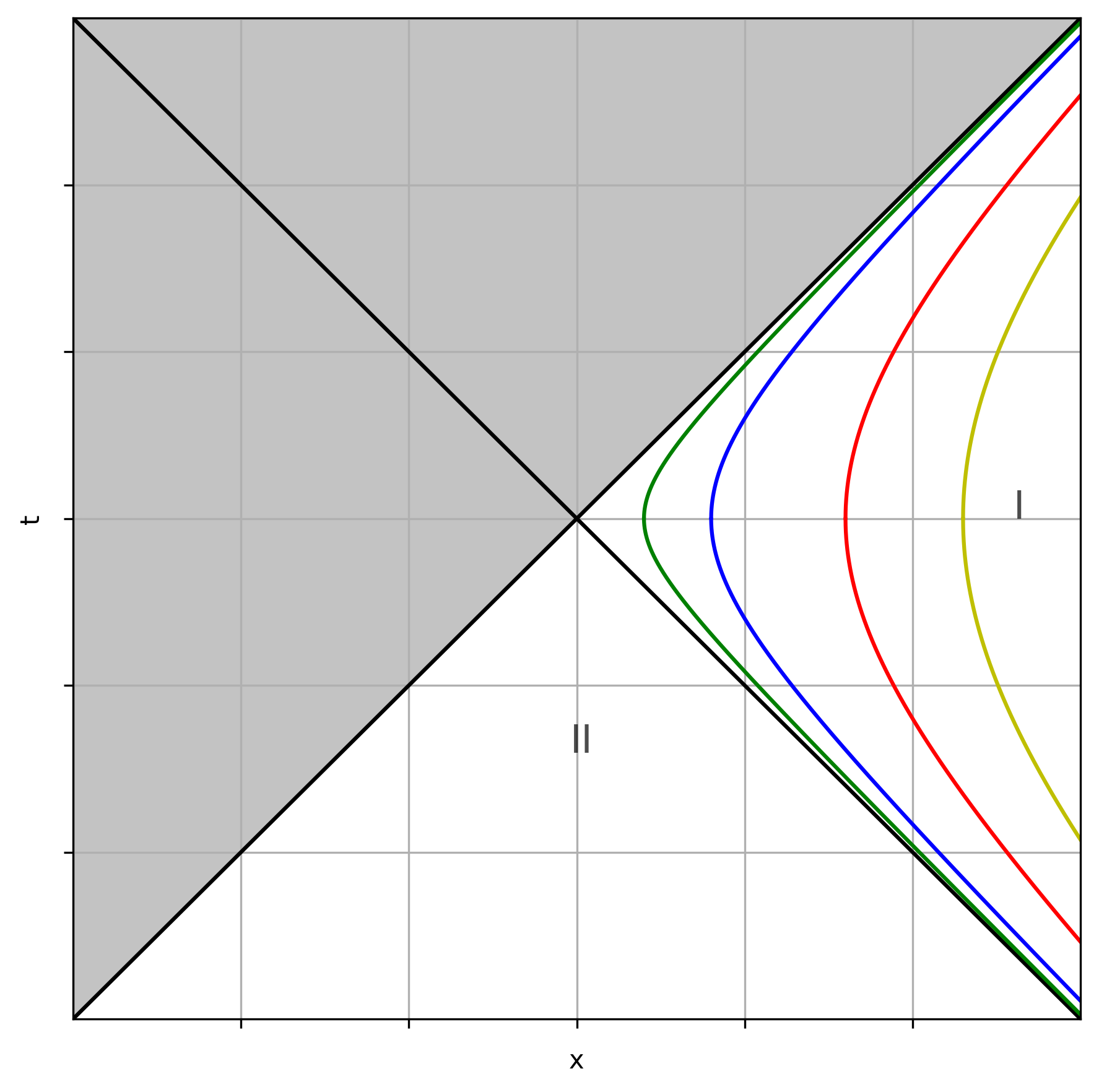}}
\vspace*{8pt}
\caption{Region I is the Rindler wedge. The shaded region is inaccessible to the $\rho =const.$ observers, hidden by the horizon $t=x$.}
\label{wedge}
\end{figure}

Consider the simplest case $+\eta =1$ in the above solutions, which leads to $x=\rho \cosh T$ and $t=\rho \sinh T$. Given that (2.12) implies $x^{2}-t^{2}=\rho ^{2}$, curves with constant $\rho $ are hyperbolas on the $(x,t)$ diagram (see Fig. \ref{wedge}). Moreover, when $\rho \rightarrow 0$ the hyperbolas asymptotically approach the light cone $x=t$. Any curve with $\rho =const.$ has slope $m>1$, indicating timelike trajectories within the local light cones.

These curves can be associated with a constant accelerated motion, which can be shown as follows: At constant $\rho >0$, the metric is $-dt^{2}+dx^{2}=-\rho ^{2}dT^{2}$, from which $dT=\rho ^{-1}d\tau$ corresponds to the rescaled proper time. Given (2.12), the four-velocity components $u^{\alpha }=dx^{\alpha }/d\tau =dx^{\alpha }/d\tau $ are $u^{0}=\cosh T$ and $u^{1}=\sinh T$, which clearly leads to $u_{\alpha}u^{\alpha }=-1$.

Differentiating once again, the four-acceleration components are $a^{0}=\rho ^{-1}\sinh T$ and $a^{1}=\rho ^{-1}\cosh T$. The magnitude is $\vec{a}^{2}=-(a^{0})^{2}+(a^{1})^{2}=\rho ^{-2}$. Then $\rho=const.$ constitutes a trajectory for an observer with uniform acceleration: smaller the value of $\rho $, greater the acceleration, approaching the light cone $x=t$ faster as time increases, although never reaching it.

The coordinate velocity is $v^{1}=u^{1}/u^{0}=dx/dt=t/x=\tanh T$, yielding the angle with respect to the $t$-axis. For a given $\rho $, $v\rightarrow 1$ (i.e. to $c$) as $T\rightarrow \infty$.

The components of the four-acceleration vector in the $(\partial _{t},\partial _{x})$
basis are $a^{\mu ^{\prime }}=(0,\rho ^{-1})$, which can be shown by using $a^{0}=\rho ^{-1}\sinh T$ and $a^{1}=\rho ^{-1}\cosh T$ in $a^{\mu }=a^{\mu ^{\prime }}\partial x^{\mu }/\partial x^{\mu^{\prime }}$ and then inverting. This makes clear that three-acceleration is also constant for constant $\rho$.

It should be evident that no particle with constant acceleration can be reached by any signal emitted within the light cone $x\leq t$. That is, a zone of unreachable events (a horizon) has been induced in spacetime. The region of Minkowski space covered by $\rho >0$ is known as \textit{the Rindler wedge} [14]. The usual (1+3)-Minkowski association is trivial: one simply adds $dy^{2}+dz^{2}$ to (2.1), which directly establishes the relationship with (1.1). See Refs. [15] and [16] for another type of transformation of accelerated motion, going from Minkowski to conformal Minkowski.

One may wonder what happened to the initial assumption about converting (2.1) to its explicit flat (1+1) Minkowski form. The answer is: in the transformation $g_{\mu \nu }=\omega ^{2}g_{\mu ^{\prime }\nu ^{\prime }}$, the information about the conformal factor is lost, since then the left-hand side of Eqs. (2.2) and (2.3) both have the factor $\omega ^{2}$, but cancellations lead to exactly the same Eqs. (2.5) and (2.6), as well as the subsequent developments.

The complementary perspective is, of course, that $-\rho ^{2}dT^{2}+d\rho^{2}$ is conformal Minkowski. For instance, the trivial case is the textbook transformation (2.12) with $+\eta =1$, while the other cases (remember $\eta=0$ is not allowed) correspond to truly curved spacetimes.

This line of thinking leads us to the next important point: we have only considered the reduced (1+1)-dimensional part. Let us explore the implications of taking full (1+3)-dimensional spherically-symmetric spacetimes.

\section{Rindler in Schwarzschild and the Spherical-Rindler Black Hole} \label{BlackHoles}

So far, we have implicitly associated Rindler-type representations with Minkowski spacetime. Now we discuss two ways in which they appear in Schwarzschild's metric: near-horizon coordinates and Kruskal-Szekeres representation. Furthermore, the former motivates to define a Spherical-Rindler black hole solution.

\subsection{From Rindler to Schwarzschild: near horizon coordinates} \label{NearHor}

Schwarzschild spacetime is generally characterized by%
\begin{equation}
ds^{2}=-\left(1-\frac{2GM}{r}\right) dt^{2}+\frac{dr^{2}}{1-\frac{2GM}{r}}+r^{2}d\Omega^{2},  \tag{3.1}
\end{equation}%
where $G$ is the gravitational constant and $M$ is the mass parameter. This form explicitly shows asymptotic flatness for $r\gg 2GM$.

This coordinate system can be associated with far-away observers. For different values of constant $r$, we have what are known as \textit{shell observers}. Since $ds^{2}=-d\tau ^{2}$, the proper time experienced by an observer at $r=const.$ compares with the one of a distant observer in the form%
\begin{equation}
dt=\frac{d\tau }{\sqrt{1-\frac{2GM}{r}}}.  \tag{3.2}
\end{equation}%

Similarly, at constant time and angles, we have%
\begin{equation}
d\varrho =\frac{dr}{\sqrt{1-\frac{2GM}{r}}},  \tag{3.3}
\end{equation}%
where $d\varrho $ is the proper radial distance in (3.1). This makes explicit the contraction of radial measurements, as seen by far-away observers.

Note that the right-hand side of (3.3) can be rewritten as%
\begin{equation}
\frac{(r-GM)dr}{\sqrt{r(r-2GM)}}+\frac{dr}{\sqrt{\left( \frac{r}{GM}%
-1\right) ^{2}-1}},  \tag{3.4}
\end{equation}%
which allows us to directly integrate (3.3), which results in%
\begin{equation}
\varrho =\sqrt{r\left( r-2GM\right) }+GM\cosh ^{-1}\left( \frac{r}{GM}%
-1\right) .  \tag{3.5}
\end{equation}%
We have set the integration constant to zero, and thus $\varrho \rightarrow 0 $ as $r\rightarrow 2GM$. The identity $\cosh ^{-1}\left[ r/(GM)-1\right] =2\sinh ^{-1}\sqrt{r/(2GM)-1}$ holds for $r>2GM$. Also, for $r$ just above the Schwarzschild radius, $sinh ^{-1}$ approximates its argument, whereas $\sqrt{r\left( r-2GM\right) }\cong \sqrt{%
2GM\left( r-2GM\right) }$. Putting it all together, the relation (3.5) near the horizon is well approximated by $\varrho =4GM\sqrt{r/(2GM)-1}$. That is
\begin{equation}
r=\frac{\varrho ^{2}}{8GM}+2GM.  \tag{3.6}
\end{equation}%
From this we obtain $(r-2GM)/r=\varrho ^{2}/\left( \varrho
^{2}+16G^{2}M^{2}\right) $, which in turn approximates $\varrho^{2}/(16G^{2}M^{2})$. This factors $dt^{2}$ in (3.1), while for $dr^{2}$ the transformation can be obtained from (3.3). Thus, for $r$ near $2GM$, the metric (3.1) is approximated by

\begin{equation}
ds^{2}=-\varrho ^{2}dT^{2}+d\varrho ^{2}+r^{2}d\Omega ^{2},  \tag{3.7}
\end{equation}%
where we have rescaled the time coordinate by $t=16G^{2}M^{2}T$.

We have shown that a Rindler transformation can also be associated with nontrivial solutions to Einstein's equations, such as Schwarzschild. The angular terms, and in particular the presence of the Schwarzschild coordinate $r$ in (3.7), prevents a complete association with flat spacetime. We can make a further approximation, since spherical symmetry allows us to consider small angles around $\theta =0$. With $r\approx 2GM\,$, the definitions $x=2GM\theta \cos \phi$ and $y=2GM\theta \sin \phi $ lead to
\begin{equation}
dx^{2}+dy^{2}=4G^{2}M^{2}\left( d\theta ^{2}+\theta ^{2}d\phi ^{2}\right) , 
\tag{3.8}
\end{equation}%
which is the first-order approximation to $r^{2}\left( d\theta ^{2}+\sin^{2}\theta d\phi ^{2}\right) $. We can then write the full metric in the Rindler form $-\varrho ^{2}dT^{2}+dx^{2}+dy^{2}+d\rho ^{2}$ (see Refs. [14][17]). All these results emphasize the highly accelerated frame in the radial direction at certain $r$. It is worth emphasizing the importance of this Rindler's approximation, given its connection with Hawking radiation, via the Unruh effect [15][18][19].

\subsection{From Schwarzschild to Kruskal-Szekeres} \label{Kruskal}

Many analytical solutions to Einstein's equations (1.3) take the form \begin{equation}
ds^{2}=-f dt^{2}+f^{-1}dr^{2}+r^{2}d\Omega ^{2},  \tag{3.9}
\end{equation}
where $f=f(r)$. These include the Schwarzschild metric, other static black hole solutions and some cosmological solutions that can be represented by static coordinates in the appropriate patch, such as De-Sitter spacetime [10].

Which metrics with $f\neq1$ in (3.9) can be transformed into the spherical-Rindler form (3.7)? One way to approach this question is to consider
\begin{equation}
-fdt^{2}+f^{-1}dr^{2}=\omega^{2}(-\rho ^{2}dT^{2}+d\rho ^{2}).  \tag{3.10}
\end{equation}

By a similar procedure to that of Sect. \ref{RindlerPrev} , one can show that the following relation between $x^{a}=(t,r)$ and $x^{a^{\prime }}=(T,\rho )$ is satisfied:
\begin{equation}
\frac{\partial t}{\partial T}=f^{-1}\rho\frac{\partial r}{\partial \rho }. 
\tag{3.11}
\end{equation}%

This equation becomes separable for $t=t(\tau)$ and $r=r(\rho)$, requiring that the differential operators be treated as total derivatives.  It is also worth noting that $f=f(\rho)$, similar to what occurs when transforming to isotropic coordinates.

Variations on the left side and the right side of (3.11) are independent and equal to a constant, say $b$. Then $dt/dT=b$, as well as $dr/f=b d\rho /\rho$, and then substitution on the left side of (3.10) lead to 
\begin{equation}
-fdt^{2}+f^{-1}dr^{2}=b ^{2}f\rho ^{-2}\left(-\rho ^{2}dT^{2}+d\rho ^{2}\right).
\tag{3.12}
\end{equation}%

The conformal factor, temporarily lost, has reappeared as $\omega^2=b ^{2}f\rho ^{-2}$. 

Separation of variables also allows to identify the transformations $t=bT$ and $\rho \propto exp({r^{\ast}/b})$, where $r^{\ast }=\int dr/f$ is the Tortoise coordinate, which arises from a simple version of conformal flatness, where the left-hand side of Eq. (3.10) is equated to $ds^2=f(-dt^2+dr^{\ast 2})$. In this formulation, the conformal factor is $f$. For the specific metric (3.1), the Tortoise coordinate takes the form $r^{\ast }=r+2GM\ln \left( r/2GM-1\right)$. This transformation maps the event horizon $r=2GM$ to $r^\ast\rightarrow-\infty$. Both the $r$ and $r^{\ast}$ coordinates explicitly demonstrate asymptotic flatness for $r\rightarrow\infty$, while $\rho\in (0,\infty)$ due to the exponential transformation.

A key difference is that in the Tortoise case the conformal factor is directly $f$, with $r^{\ast}=r^{\ast}(r)$ and $t$ unchanged. By contrast, here $\rho=\rho(r)$, while $T\propto t$. The implications of this are significant, as a transformation of the type (2.12) must still be applied, rendering the result non-trivial.

For Schwarschild's case $\rho ^{2}=e^{r/2M}\left( r/2M-1\right)$, given $b =4M$. Equation (3.12) then leads to
\begin{equation}
ds^{2}=\frac{32M^{3}}{r}e^{-\frac{r}{2M}}(-\rho ^{2}dT^{2}+d\rho
^{2})+r^{2}d\Omega ^{2}.  \tag{3.13}
\end{equation}%
By performing the custom hyperbolic transformation $\xi =\rho\sinh T$ and $R=\rho \cosh T$, we recognize the custom Kruskal-Szekeres representation.

Thus, with the appropriate translation, all this allows us to reinterpret Fig. \ref{wedge} as the Kruskal-Szekeres diagram. Region I corresponds to the black hole's exterior, with the other regions constructed by means of maximal extension [14][20][21].

\subsection{Spherical-Rindler Black Hole type solution} \label{spherical-Rindler-BH}

According to (3.12), we obtain directly a Rindler representation for the two first terms of (3.9) when $\rho ^{2}=b ^{2}f$, which implies --see just above Eq. (3.12)-- the differential relation $bdr= \rho d\rho$, with general solution  
\begin{equation}
\rho ^{2}=b ^{2}f=2b r+c,  \tag{3.14}
\end{equation}%
where $c$ is another constant. This represents a broad class of spherically symmetric metrics with Rindler representation, depending on the values of the constant parameters. By taking a glimpse at Eq. (3.6) one recognizes that it belongs to this class of transformations. That is, if we extend the approximation of Sect. \ref{NearHor} to be a global metric \textit{per se}, we obtain a spherically symmetric Rindler metric.

By choosing $b=2r_{h}$ and $c=-4r_{h}^2$ in (3.14), then (3.9) adopts the form

\begin{equation}
ds^{2}=-\left( \frac{r}{r_{h}}-1\right) dt^{2}+\frac{dr^{2}}{\frac{r}{r_{h}}-1}+r^{2}d\Omega ^{2}.  \tag{3.15}
\end{equation}

This metric, where $g_{11}\rightarrow \infty $ as $r\rightarrow r_{h}$, describes a black hole solution. The region $r>r_{h}$ constitutes the causal patch where it is is timelike and $\partial _{r}$ is spacelike,  and the geometry approximates Schwarzschild spacetime near the horizon $r=r_{h}$.

It is instructive to contrast some features of this metric with those of the usual Rindler-Minkowski correspondence and the Schwarzschild case. First, since our metric (3.15) is independent of $t$ and $\phi$, the vector fields $\partial _{t}$ and $\partial _{\phi }$ are Killing vectors, a property shared with both Minkowski and Schwarzschild spacetimes. In Minkowski spacetime, the independence of the terms $-\rho ^{2}dT^{2}+d\rho ^{2}$ on the Rindler time $T$ (derived from the transformations $t=\rho \sinh T$ and $x=\rho \cosh T$) generates the boost Killing vector $\partial _{T}=x\partial_{t}+t\partial _{x}$ . There are three independent boost directions, and along with the four translational and three rotational Killing vectors, complete the set of ten required for a maximally symmetric spacetime. In contrast, for the metric (3.15), the Killing vector  $\partial _{T}=(2r_{h})^{-1}\partial _{t}$ is not associated with any additional symmetry. Our solution, like the Schwarzschild solution, possesses only four Killing vectors: one temporal and the three rotational ones.

A second key distinction is that while we have highlighted the relationship of this metric with near-horizon coordinates, unlike the Schwarzschild metric (3.1), it is not asymptotically flat, behaving as $r/r_h$ whenever $r\rightarrow \infty$.

To obtain properties of this spacetime, one can start with Christoffel symbols
\begin{equation}
\Gamma _{\alpha \beta }^{\mu }=\frac{1}{2}g^{\mu \lambda }(\partial _{\alpha
}g_{\lambda \beta }+\partial _{\beta }g_{\alpha \lambda }-\partial _{\lambda
}g_{\alpha \beta }),  \tag{3.16}
\end{equation}
and then use the dynamical equation (1.3) or the geodesic equation. Instead, we consider the kinematic effects using the Killing vectors $T^{\mu }=\delta _{0}^{\mu }$ and $\Phi ^{\mu }=\delta_{3}^{\mu }$. They imply $u_{0}=g_{00}dx^0/d\tau$ and $u_{3}=g_{33}dx^3/d\tau$ are conserved [14][22]. That is,
\begin{equation}
\left( \frac{r}{r_{h}}-1\right) \dot{t}=e  \tag{3.17}
\end{equation}%
and%
\begin{equation}
r^{2}\sin ^{2}\theta ~\dot{\phi}=l,  \tag{3.18}
\end{equation}%
where $e$ and $l$ are constants. Overdots indicate derivation respect to proper time $\tau$.

Now consider massive particles, with four-velocity normalized as $u^{\alpha }u_{\alpha }=-1$. This condition is equivalent to%
\begin{equation}
-\left( \frac{r}{r_{h}}-1\right) \dot{t}^{2}+\frac{\dot{r}^{2}}{\frac{r}{%
r_{h}}-1}+r^{2}\dot{\theta}^{2}+r^{2}\sin ^{2}\theta ~\dot{\phi}^{2}=-1. 
\tag{3.19}
\end{equation}%
 Equation (3.18) and spherical symmetry allows to consider orbits in the plane $\theta =\pi /2$. Substituting the expressions for $\dot{t}$ and $\dot{\phi}$ in Eq. (3.19) leads to
\begin{equation}
\dot{r}^{2}+\left( 1+\frac{l^{2}}{r^{2}}\right) \left( \frac{r}{r_{h}}%
-1\right) =e^{2},  \tag{3.20}
\end{equation}%
while $\dot{\phi}=l/r^{2}$. This radial equation for our metric can be compared to the Schwarzschild case, where the last parenthesis would be $\left( 1-2M/r\right)$ instead.

Equation (3.20) can be rearranged as follows:
\begin{equation}
\frac{\dot{r}^{2}}{2}+\frac{l^{2}}{2}\left( \frac{1}{r_{h}r}-\frac{1}{r^{2}}+%
\frac{r}{l^{2}r_{h}}\right) =\frac{e^{2}+1}{2}.  \tag{3.21}
\end{equation}%

At the left, the first term represents kinetic energy, and the remaining terms are the effective potential energy $U_{eff}$. In Schwarzschild, it would consist of three terms: a Newtonian gravitational term proportional to $-r^{-1}$, a centrifugal barrier proportional to $r^{-2}$, and a relativistic correction proportional to $-r^{-3}$ [22].

The effective potential in (3.21) possesses different features. It has  terms with $r^{-1}$ and $r^{-2}$ dependence, but with opposite signs to Schwarzschild's case. Since $r>r_{h}$, the first term $l^2/(2r_{h}r)$ dominates the second one, $-l^2/(2r^2)$, creating a net radial repulsive fictitious force for $l\neq 0$. The linear term in $r$, which dominates when $r\gg r_{h}$, is the only term for purely radial trajectories. In the Newtonian association, this linear term corresponds to a radial force of constant magnitude, a result consistent with the Rindler and uniform acceleration association.

Circular orbits are obtained by setting $dU_{eff}/dr=0$.This leads to a cubic equation for the radius, $r^3-l^2(r-2r_h)=0$, which has physical solutions only for $r>2r_h$ [23]. Three possibilities exist. When $l^2<27r_h^2$, the cubic yields only a negative real solution, i.e. no circular orbits exist. The critical case $l^2=27r_h^2$ results in a single positive solution at $r=3r_h$. This marks the transition to the third case, $l^2>27r_h^2$, which presents two solutions in the region of interest. The lower-radius solution indicates an unstable circular orbit whose radius decreases from $r=3r_h$ and asymptotically approaches $r=2r_h$ as $l^2$ increases. This lower limit coincides with the radius of the photon sphere. The second solution corresponds to a stable circular orbit (a minimum of $U_{eff}$). The lower value $r=3r_h$ constitutes the \textit{innermost stable circular orbit (ISCO)}. The radius for other stable circular orbits increases unbounded when $l^2$ increases.

\begin{figure}[ph]
\centerline{\includegraphics[width=3.0in]{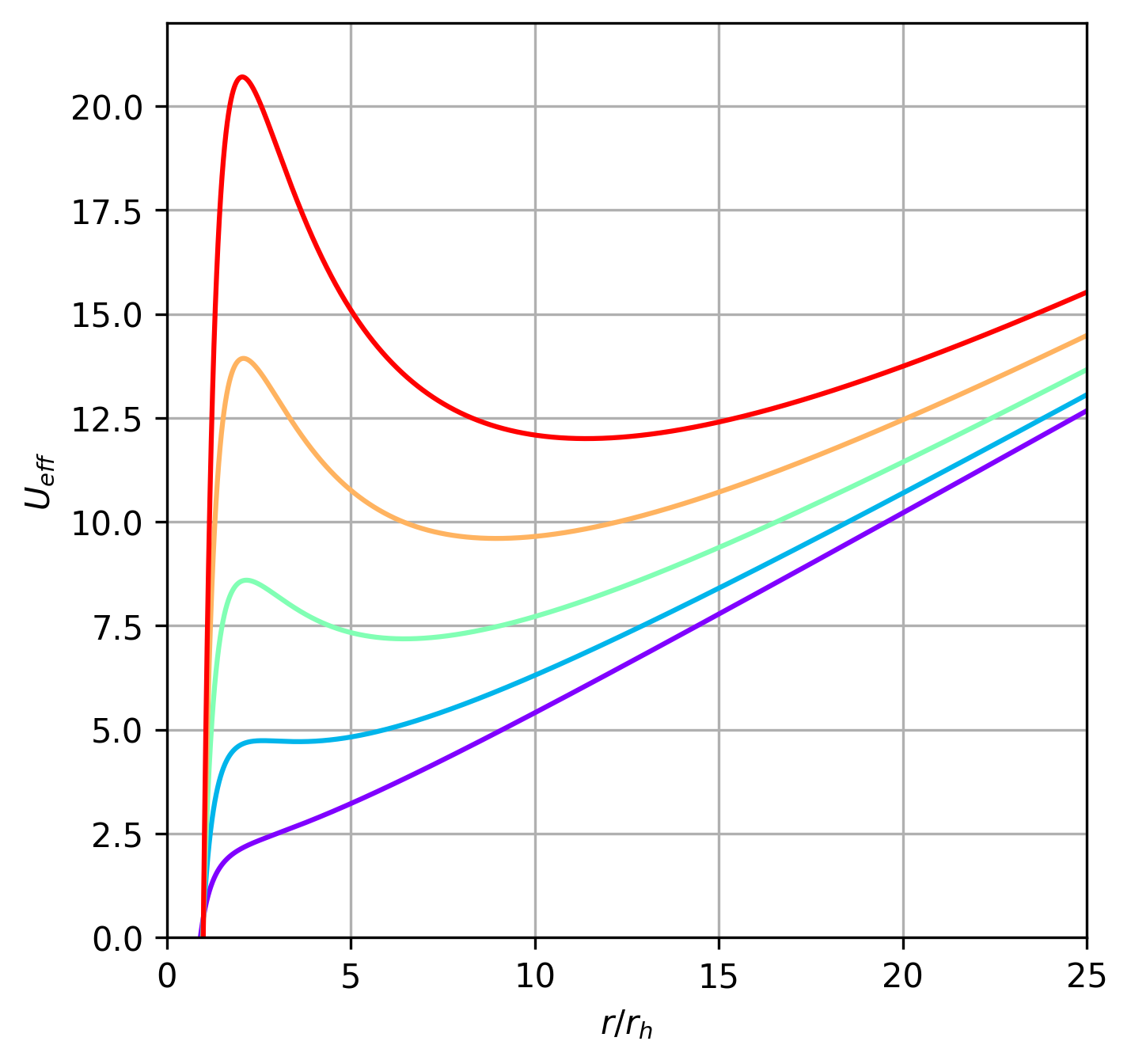}}
\vspace*{8pt}
\caption{Effective potentials for spacetime (3.15). The lower curve is the only one with $l^2<27r_h^2$.}
\label{Potential}
\end{figure}

All of these features are depicted in Fig. \ref{Potential}. Bounded orbits occur when the energy parameter $(e^{2}+1)/2$ lies between the local minima and maxima of $U_{eff}$.

For massless particles, in (3.20) one has $l^2/r^2$ in place of the $(1+l^2/r^2)$ factor. In our actual case, $r=3r_h$ for the ISCO and $r=2r_h$ for the photon sphere. In Schwarzschild spacetime they are $r=3r_s$ and $r=3/2r_s$, respectively [22].

While this spacetime has other interesting properties, it is worth noting that if we embed its spatial sections in Euclidean space, it can only be represented for the region $r_h<r<2r_h$. In contrast, Schwarzschild spacetime (3.1) is represented by Flamm's paraboloid all the way to $r\rightarrow \infty$ [24].

In sum, this black hole spacetime exhibits a curious 1-2-3 $r_h$-effect. The radius $r=r_h$ marks the event horizon. The region $r_h\leq r \leq 2r_h$ is the only part that can be embedded in Euclidean space, and it also contains the photon sphere. The region $2r_h\leq r  \leq 3r_h$ admits unstable circular orbits for massive test particles. Also, the radius $r=3r_h$ is the location of the innermost stable circular orbit (ISCO), and the region $r>3r_h$ presents stable circular orbits for $l^2>27r_h^2$.

\section{Rindler representations in cosmology} \label{cosmology}

In this section we discuss two cosmological spacetimes that can be related to Rindler transformations: a basic and well-known FLRW solution and a spherical-Rindler metric cosmological solution.

\subsection{Milne universe} \label{Milne}

The general FLRW metric takes the form%
\begin{equation}
ds^{2}=-dT^{2}+a^{2}\left( \frac{dR^{2}}{1-kR^{2}}+R^{2}d\Omega ^{2}\right) ,
\tag{4.1}
\end{equation}%
where $a=a(T)$ is the scale parameter that characterizes the expansion of the universe. The part in parentheses describes an isotropic and homogeneous spatial section at $T=const.$, where $k$ may be set to $1$, $0$ or $-1$. Consistent with this, by assuming a perfect fluid energy-momentum tensor, the $00$ component of Einstein equation (1.3) results in the Friedmann equation%
\begin{equation}
\frac{\dot{a}^{2}+k}{a^{2}}=\frac{8\pi G}{3}\varrho _{m}+\frac{\Lambda }{3},
\tag{4.2}
\end{equation}%
where $\varrho _{m}$ is the energy density. The case $\varrho _{m}=k=\Lambda=0$ yields directly Minkowski, while a less trivial case results when $\varrho _{m}=\Lambda =0$, $k=-1$. We then have $\dot{a}^{2}-1=0$, with $a=T$ as a suitable solution. (4.1) now takes the explicit form%
\begin{equation}
ds^{2}=-dT^{2}+T^{2}\left( \frac{dR^{2}}{1+R^{2}}+R^{2}d\Omega ^{2}\right) .
\tag{4.3}
\end{equation}%
This {\it Milne universe} (see Ref. [10] and references therein).
This metric describes an expanding universe with hyperbolic topology, a spacetime dominated by curvature. Given that $\varrho _{m}$ and $\Lambda $ are both zero, it is reasonable to question whether the spacetime is truly just Minkowski space.

To demonstrate this, we can first examine the hyperbolic character of the expanding spatial sections. This manifests itself by performing the transformation $d\chi =dR/(1+R^{2})$, with solution $R=\sinh \chi$. With this transformation to the hyperbolic angle $\chi$, the problem becomes equivalent to showing that a coordinate transformation exists from $-dT^{2}+T^{2}d\chi ^{2}$ to $-d\zeta ^{2}+d\xi ^{2}$, given $r=TR$. At this point, the necessary transformation is a familiar one:%
\begin{equation}
\zeta =T\cosh \chi ,  \tag{4.4}
\end{equation}%
and 
\begin{equation}
\xi =T\sinh \chi=TR.  \tag{4.5}
\end{equation}

 We need to include the angular terms to secure that Milne universe is flat spacetime. By (4.5), the solid angle $d\Omega^2$ in (4.3) has factor $\xi^2$, rendering the (1+3) metric (4.3) into $-d\zeta ^2+d\xi^2+\xi^2d\Omega^2$. As $\zeta ^{2}-\xi ^{2}=T^{2}$ and $\xi =\zeta \tanh \chi$ hold, we name {\it Milne wedge} the region $T>0$ of (4D) Minkowski space that it represents. For considering more relationships between Rindler, Milne and Schwarschild, see [25].

\subsection{Spherically symmetric static cosmological solution} \label{spherical-RindlerCosmo}

Some cosmological spacetimes like De Sitter, Anti-de-Sitter (AdS), among a few others, can be expressed in the static form (3.9), with $f=f(r)$ [7][10].

Another cosmological solution arises by setting $b=-2l$ and $c=4l^2$ in Eq. (3.14), for which the metric (3.9) takes the form
\begin{equation}
ds^{2}=-\left( 1-\frac{r}{l}\right) dt^{2}+\frac{dr^{2}}{1-\frac{r}{l}}+r^{2}d\Omega ^{2}.  \tag{4.6}
\end{equation}

That $r=l$ is a cosmological horizon can be seen by the fact that the vectors $\partial _t$ and $\partial _r$ are respectively timelike and spacelike for $r<l$. This metric is isotropic, and is locally Minkowski for $r\rightarrow 0$, a property shared with other cosmological solutions. However, it is not homogeneous.

We are interested in the causal patch $r<l$. From Christoffel's symbols (3.16) we calculate the components of the Ricci tensor by%
\begin{equation}
R_{\mu \nu }=\partial _{\alpha }\Gamma _{\mu
\nu }^{\alpha }-\partial _{\nu }\Gamma _{\mu \alpha }^{\alpha }+\Gamma
_{\alpha \lambda }^{\alpha }\Gamma _{\mu \nu }^{\lambda }-\Gamma _{\nu
\lambda }^{\alpha }\Gamma _{\mu \alpha }^{\lambda }.  \tag{4.7}
\end{equation}%
The non-zero components of $R_{\mu \nu }$, and the scalar
curvature $R=g^{\mu \nu }R_{\mu \nu }$, are%
\begin{equation}
\begin{tabular}{l}
$R_{00}=-\frac{1}{l^{2}r}\left( l-r\right) ,$ \\ 
$R_{11}=\frac{1}{r(l-r)},$ \\ 
$R_{22}=\frac{2r}{l}$, $R_{33}=\sin ^{2}\theta ~R_{22},$ \\ 
$R=\frac{6}{lr}.$%
\end{tabular}
\tag{4.8}
\end{equation}

The components of the energy-momentum tensor can be calculated using
Einstein's equations (1.3). With $\Lambda =0$, they are expressed as%
\begin{equation}
\begin{tabular}{l}
$8\pi T_{00}=\frac{2}{l^{2}}\left( \frac{l}{r}-1\right) ,$ \\ 
$8\pi T_{11}=-\frac{2}{r^{2}\left( \frac{l}{r}-1\right) },$ \\ 
$8\pi T_{22}=-\frac{r}{l}$, $8\pi T_{33}=\sin ^{2}\theta ~T_{22}$ \\ 
$8\pi T=-\frac{6}{lr}.$%
\end{tabular}
\tag{4.9}
\end{equation}

Compare the results with those of the following metric:

\begin{equation}
ds^{2}=-\left( 1-\frac{r^{2}}{l^{2}}\right) dt^{2}+\frac{dr^{2}}{1-\frac{r^{2}}{l^{2}} }+r^{2}d\Omega ^{2}.  \tag{4.10}
\end{equation}%
This form is often associated with De Sitter spacetime, for which the usual FLRW representation is obtained by setting $\varrho _{m}=k=0$ in (4.1) and (4.2) [5][10].

Computing the components of the Ricci tensor, we obtain $R_{00}=-3\left(
l^{2}-r^{2}\right) /l^{4}$, $R_{11}=3\left( l^{2}-r^{2}\right) $, $%
R_{22}=R_{33}\sin ^{-2}\theta =3r^{2}/l^{2}$ and $R=12/l^{2}$. Again, by means of (1.3), we obtain
\begin{equation}
\begin{tabular}{l}
$8\pi T_{00}=\frac{3}{l^{2}}\left( 1-\frac{r^{2}}{l^{2}}\right) ,$ \\ 
$8\pi T_{11}=-\frac{3}{l^{2}\left( 1-\frac{r^{2}}{l^{2}}\right) },$ \\ 
$8\pi T_{22}=-\frac{3r^{2}}{l^{2}}$, $T_{33}=\sin ^{2}\theta $~$T_{22},$ \\ 
$8\pi T=-\frac{12}{l^{2}}.$%
\end{tabular}
\tag{4.11}
\end{equation}

Compare the metrics (4.6) and (4.10), as well as the results (4.9) and (4.11): some features are similar, such as $T_{00}\rightarrow 0$ when $r\rightarrow l$,
the cosmological horizon. However, in De Sitter spacetime, the trace of the energy-momentum tensor is constant: it is a maximally symmetric spacetime. Notice that $l^{2}=3/\Lambda $ produces the connection between the static version (4.10) and the explicit FLRW solution (4.1).

An important point: all of these results refer to the coordinate basis associated with the metric form (4.6), i.e. with respect to $\partial _{\alpha }$, considering $x^{\mu }=(t,r,\theta ,\phi)$. It is also worth noting that (4.1) can only be transformed into the static
form (4.10) in a few possibilities, all with $\varrho _{m}=0$ in the Friedmann equation (4.2) [11]. Some examples: $k=-1$ and $\Lambda =0$ yields the aforementioned Milne solution (4.3), while ($k=0$, $\Lambda >0$) and ($k=-1$, $\Lambda <0$) would correspond to De Sitter and Anti De Sitter solutions, respectively. For the
De Sitter solution, (4.2) gives $a=\exp (\sqrt{\Lambda /3}T)$. Now, while in general the generic FLRW cosmological models (4.1) are maximally symmetric
(isotropic and homogeneous) in spatial sections, De Sitter and Anti De Sitter share the additional property of being maximally symmetric in spacetime. This is reflected in the relationship $R=-8\pi T=12/l^{2}$.

\section{Summary and final comments}

Given the covariant nature of general relativity (GR), the role of coordinate transformations is often considered a secondary issue. However, when analyzing specific models of the universe, the way a given spacetime is presented is crucial. For example, while Schwarzschild spacetime is usually derived from a static, asymptotically flat radial ansatz, other representations allow for more properties to be extracted. For instance, Painlevé-Gullstrand coordinates can be used to account for measurements from freely falling observers, while Eddington-Finkelstein or Lema\^{\i}tre coordinates allow for the crossing of the event horizon. Furthermore, Kruskal-Szekeres and conformal representations capture the causal structure of the spacetime and allow extending it [1][14]. Similar arguments apply to cosmological models.

In this article we have reviewed Rindler-type transformations that can be associated with various solutions in GR. Starting from flat spacetime, we obtained the wide class of transformations (2.12) and (2.13). These preserve the common association with accelerated frames at constant $\rho$. Also, the process described in Sect. \ref{RindlerPrev} allowed us to obtain the set of Rindler-type cyclic transformations (2.14)-(2.16), which represents a compact region of Minkowski spacetime.

We then extended the application of Rindler-type transformations to the realm of black holes and cosmology. We provided an alternative, Rindler-based derivation of the Kruskal-Szekeres representation, bypassing the typical textboook sequence Tortoise $\rightarrow$ Eddington–Finkelstein $\rightarrow$ Kruskal-Szekeres. Also, in Sect. \ref{spherical-Rindler-BH} we went deeper into the properties of the black hole spacetime described by metric (3.15), a Spherical-Rindler black hole solution.

We also found Rindler representations in cosmology. The Milne FLRW solution can be viewed as a Rindler-type transformation of Minkowski spacetime,. As with the black hole solution of Sect. \ref{spherical-RindlerCosmo}, there is another spherical-Rindler-type solution, with a cosmological horizon, that can also be represented by (3.7). Here we compared some aspects with the De Sitter metric, which similarly allows for a static representation inside its cosmological horizon.

It is worth to notice that our Spherical-Rindler black hole and cosmological solutions have been explored in distinct contexts, as an approximation or originated from modified gravity (see Refs. [25][26][27]). Our approach emphasizes the geometrical origin and the Rindler connection.

Some open questions, such as more properties of the cosmological solution, possible applications of these metrics, the curious properties of flat embedding, among others, are beyond the scope of this article and are left for future study.

\section*{Acknowledgments}
The author thanks A. Sandoval-Rodr\'{\i}guez and E.d.J. Le\'{\o}n-Mart\'{\i}nez for helpful comments.
\\


\begin{thebibliography}{0}

\bibitem{MTW} C. W. Misner, K. S. Thorne and J. A. Wheeler, Gravitation (W. H. Freeman and Company, San Francisco, 1973).

\bibitem{Dieks} D. Diek, Another Look at General Covariance and the Equivalence of Reference Systems, {\it Stud. Hist. Philos. Mod. Phys.} {\bf 37}(1) (2003) 174.

\bibitem{deSitter} W. de Sitter, On Einstein's Theory of Gravitation and its Astronomical Consequences. First Paper, {\it MNRAS} {\bf 76}(9) (1916) 699.

\bibitem{Lemaitre} G. Lema\^{\i}tre, L'Univers en expansion, {\it Annales Soc. Sci. Bruxelles} {\bf A53} (1933) 51.

\bibitem{Tolman} R. C. Tolman, Relativity, Thermodynamics and Cosmology (Clarendon Press, Oxford, London, 1934).

\bibitem{Tanatarov} I. V. Tanatarov and O. B. Zaslavskii, Dirty rotating black holes: Regularity conditions on stationary horizons, {\it Phys. Rev. D} {\bf 86}(4) (2012) 044019.

\bibitem{Mitra} A. Mitra, When can an "Expanding Universe" look "Static" and vice versa: A comprehensive study, {\it Int. J. Mod. Phys. D} {\bf 24}(5) (2015) 1550032.

\bibitem{Zaslavskii} O.B. Zaslavskii,  On regular frames near rotating black holes, {\it Gen. Rel. Grav.} {\bf 50}(10) (2018) 123.

\bibitem{Visser} M. Visser, Efficient computation of null affine parameters, {\it Universe} {\bf 9}(12) (2023) 521.

\bibitem{Leon1} E. A. Le\'{o}n and A. Sandoval-Rodr\'{\i}guez, Symmetry Transformations in Cosmological and Black Hole Analytical Solutions, {\it Symmetry} {\bf 16}(4) (2024) 394.

\bibitem{Leon2} E. A. Le\'{o}n, J. A. Nieto, A. Sandoval-Rodr\'{\i}guez and B. Mart\'{\i}nez-Olivas, Beyond Schwarzschild: new pulsating coordinates for spherically symmetric metrics, {\it Gen. Rel. Grav.} {\bf 56}(3) (2024) 35.

\bibitem{Rindler1} W. Rindler, Hyperbolic Motion in Curved Space Time, {\it Phys. Rev.} {\bf 119}(6) (1960) 2082.

\bibitem{Rindler2} W. Rindler, Kruskal Space and the Uniformly Accelerated Frame, {\it Phys. Rev.} {\bf 34}(12) (1966) 1174.

\bibitem{Dray} T. Dray, Differential forms and the geometry of general
relativity (CRC Press, Boca Raton FL, 2015).

\bibitem{Socolovsky} M. Socolovsky, {\it Rindler Space and Unruh Effect}, (2013); arXiv:1304.2833 [gr-qc.

\bibitem{Possel} M. P{\"o}ssel, Counterintuitive properties of relativistic
relative motion for accelerated observers, {\it Am. J. Phys.} {\bf 92}(12) (2024) 957.

\bibitem{Susskind} L. Susskind and J. Lindesay, An Introduction To Black Holes, Information And The String Theory Revolution: The Holographic Universe (World Scientific, Hackensack USA, 2005).

\bibitem{Davies} P. C. W. Davies, Scalar particle production in Schwarzschild and Rindler metrics, {\it J. Phys. A: Math. Gen. } {\bf 8}(4) (1975) 609.

\bibitem{Unruh} R. Schutzhold and W. G. Unruh, On the origin of the particles in black hole evaporation, {\it Phys. Rev. D} {\bf 178} (2008) 041504.

\bibitem{Kruskal} M. D. Kruskal, Maximal Extension of Schwarzschild Metric, {\it Phys. Rev.} {\bf 119}(5) (1960) 1743.

\bibitem{Szekeres} G. Szekeres, On the Singularities of a Riemannian Manifold, {\it Publ. Math. Debrecen} {\bf 7} (1960) 285.

\bibitem{Carroll} S. M. Carroll, Spacetime and geometry: An introduction to
general relativity (Addison-Wesley, San Francisco CA, 2004).

\bibitem{McKelvey} J. P. McKelvey, Simple transcendental expressions for the roots of cubic equations, {\it Am. J. Phys.} {\bf 53}(3) (1984) 269.

\bibitem{Marolf} D. Marolf, Spacetime Embedding Diagrams for Black Holes, {\it Gen. Rel. Grav.} {\bf 31}(6) (1999) 919.

\bibitem{Culetu} H. Culetu, The Milne spacetime and the hadronic Rindler horizon, {\it Int. J. Mod. Phys. D} {\bf 19}(8) (2010) 1379.

\bibitem{Grumiller} D. Grumiller and F. Preis, The Milne spacetime and the hadronic Rindler horizon, {\it Int. J. Mod. Phys. D} {\bf 20}(14) (2011) 2761.

\bibitem{Casadio} R. Casadio, A. Kamenshchik and J. Ovalle, Cosmology from Schwarzschild black hole revisited, {\it Phys. Rev. D} {\bf 110}(4) (2024) 044001.

\end{thebibliography}
\end{document}